# Crime Patterns in Los Angeles County Before and After Covid19

Rubab Hussain, Rigo Vargas, Hieu Hughes Le-Au, Will Gass, Melissa Fenn, Briseyda Serna-Marquez, Jongwook Woo
Department of Information Systems, California State University Los Angeles
{rhussai5, rvarga30, hleau, wgass, mfenn, bsernam, jwoo5}@calstatela.edu

**Abstract:** The objective of our research is to present the change in crime rates in Los Angeles post-Covid19. Using data analysis with Geo-Mapping, bubbles, Marimekko, and a time series charts, we can illustrate which areas have the largest crime rate, and how it has changed. Through regression modeling, we can interpret which locations may also have a correlation to crime versus income, race, type of crime, and gender. The story will help to uncover whether the areas associated with crime are due to demographic or income variance. In showing the details of crimes in Los Angeles along with the factors at play we hope to see a compelling relationship between crime rates and recent events from 2020 to the present, along with changes in crime type trends during these periods. We use Excel to clean the data for SAP SAC to model effectively, as well as resources from other studies a comparison.

## 1. Introduction

We began our research with the hypothesis of seeing a large drop in crimes during/post Covid19, due to a global pandemic and stay-at-home orders, crime rates should drop. This paper uses SAP Analytics Cloud to process and visualize the historical data of crimes in Los Angeles. The dataset was retrieved from LaCity.org and consisted of information type of crime, region, sex, weapon, status, and age. We have two data sources that will be compared, pre covid19 data and then post covid19 data.

## 2. Related Work

Around the world, the everyday lifestyles of people changed due to the pandemic, including the patterns of those who commit crimes. Several studies have been done to examine the effects the pandemic had on crime rates. Reports from many developed nations have shown an overall general decrease in crime. A study done by Langton, Dixon, and Farrell [3], crime data from March 2020 to August 2020 was used to analyze the crime trends during the first 6 months of the nationwide lockdown in England and Wales. The study provided further iterations that national and local lockdowns caused declines in most recorded crimes. Jason Yeung [4] compared the crime rates in Boston, Massachusetts from the pre-Covid-19 years to crime rates in the year 2020. An overall glance at the crime rates would show that the pandemic had a positive effect on crime, in that crime rates decreased. However, by using the crime data analysis, Yeon was able to distinguish the types of crimes that decreased and the crimes that had a spike during the stay-at-home orders. An interesting finding was the significant increase of incidents involving assault or disputes. In a similar study done by Felson, Jiang, Xu [2], the changes in burglaries during the first month of the pandemic [2] in the different areas of Detroit, Michigan were analyzed. This study was able to provide suggestions to law enforcement and security on the shifting of resources and monitoring of vulnerable businesses.

## 3. Specifications

The data we collected was supplied through the LA CITY public data portal [5]. We downloaded the Crime data from 2010-2019 and for 2020-Present [6, 7]. We filtered the data set in Excel and keep the Crime Data for 2018-2019 compared with 2020-to present to maintain consistency with a 1-year analysis. We also downloaded LA county COVID cases from the LA CITY portal [8].

The datasets were cleaned in Excel to remove empty values and maintain critical information, such as crime description, weapons, area, longitude, and latitude, etc. With the COVID comparisons, we combined the cleaned 2018-2019 dataset with the 2020-present dataset to overlay a model to visualize the changes.

Table 1. Data Specifications

| Crime Data from 2010 to 2019 | 155.2 MB |
|---|---|
| Crime Data from 2020 to Present | 151.9 MB |
| LA County COVID Cases | 76 KB |

## 4. Implementation Flowchart

The raw data sets were retrieved from the LA City Data website and contained crime information from different areas of Los Angeles. The flowchart below illustrates the different steps in the data manipulation process. The data used was obtained through multiple CSV files. The data files were then uploaded to SAP to further clean and create stories with models. Part of the cleaning process in SAP includes setting the right dimensions, measures, properties, and descriptions.

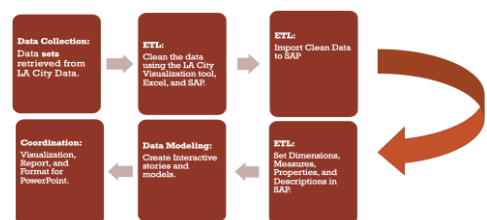

Figure 1. Implementation Flowchart

## 5. Data Engineering

The first data set initially included data from 2010 through 2019 so we filtered the data to 2018 through 2019 using the visualization tool in the LA City Data website. The remaining datasets were retrieved without any manipulation. The datasets were then further cleaned out using excel to remove "0" latitude and longitude locations. The CSV files were initially cleaned using the LA City Data website's visualization tool to filter and remove any data prior to 2018. Extra columns were added to the crime datasets to include the "area" Latitude and Longitudes, Address, State, and Zip

Code. This data was added to visualize the number of crimes each police division was handling. For the COVID comparison we combined the cleaned 2018-2019 dataset with the 2020-present dataset to overlay a model and visualize changes. Since the goal was to visualize the number of crimes by area, a column was added to the combined dataset titled "Crime #" to use as our measure in SAP. The clean CSV files were then uploaded to SAP for final cleaning and modeling. The next steps were to set the proper measures, dimensions, and formatting to model the data. First, the column labeled "Date OCC" was changed from a measure to the proper date dimension formatting. Next, we concatenated the "Month" and "TIME OCC" columns to create a timestamp. Finally, we set all other columns as dimensions and set our geo coordinates by area. The dataset was then ready to be used to create a story in SAP.

## 6. Data Analysis and Visualization

After data engineering and determining what data would best describe our analysis, a story and predictive model were created in SAP Analytics Cloud that provided a visual representation of the number of crimes per area, total volume per crime type/location, a variance analysis comparing crime before/after COVID, the relationship of new COVID cases with crime rates, and a regression predictive model to determine the influencers of crime.

### 6.1 Total Crime by Area Chart (2018 – Present)

The first visualization (Figure 2) created in SAP was using a bar chart to illustrate the concentration of crime by area sorted from highest to lowest. This quickly gives a brief overview of the areas most affected by crime. The top three areas are 77th Street (57,429), Central (54,035), and Southwest (52,308).

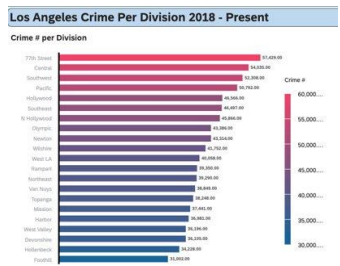

Figure 2. Total Crime by Area Bar Chart

### 6.2 Total Crime by Area Map (2018 – Present)

Figure 3 Geo-Map highlights which areas have the highest concentration of crimes in Los Angeles. The bubble color and bubble size represent the total number of crimes per area. 77th Street, which is located near the cities of Watts and South Central, is the area with the highest concentration of crime.

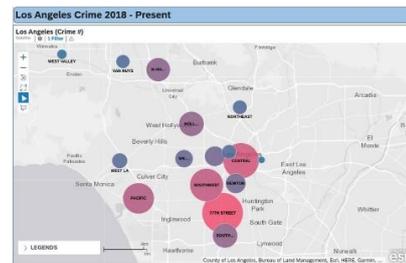

Figure 3. Total Crime by Area Geo-Map

### 6.3 Crime Before and After COVID

The bar chart below (Figure 4) illustrates the variance in crime from 2018 – Present. It gives us an understanding of crime levels before the COVID-19 pandemic (2018 – 2019) and crime after COVID-19 (2020 – Present). Prior to the pandemic, crime was already steadily shifting downward by -4.89%. At the start of the pandemic (2020) crime took a more notable dip of -9.40%. As we can see from Figure 4 below, crime has started to pick back up to pre-pandemic levels. This can be attributed to lesser restrictions, increased mobility within the population, and the lockdown mandates removed.

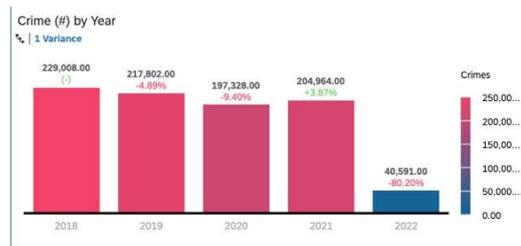

Figure 4. Crime Before and After COVID

### 6.4 New COVID Cases in LA County vs Crime

In our analysis we decided to compare the number of new COVID cases in LA County to the number of crimes committed during the pandemic to understand if the pandemic had any plausible relationship to crime rates. Figure 5 below represents the number of new COVID cases in LA county per month. Figure 6 represents all crime by month during the pandemic. To our surprise, an increase in COVID cases had no direct correlation to crime. From looking at Figure 5 below, all crime remained stagnant during the pandemic. By digging a little deeper into our analysis, we decided to narrow our focus to crimes committed by juveniles. With lockdowns in place and schools being shut down, juvenile crime could be directly affected by the pandemic. As seen in Figure 7, juvenile crime substantially decreased and fluctuated with the pandemic. When comparing Figure 5 and Figure 7, we can see a more plausible correlation. As the number of new COVID cases increased in LA County, the number of crimes committed by juveniles decreased.

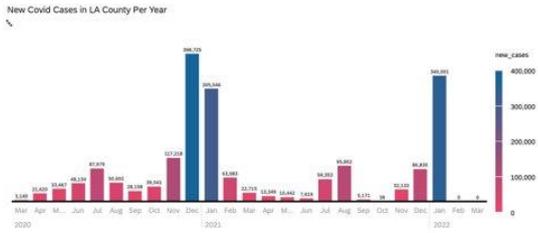

Figure 5. New COVID Cases in LA County (2020 – Present)

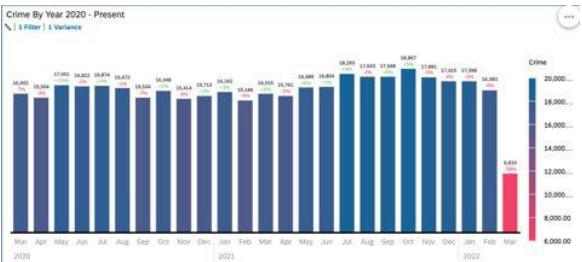

Figure 6. All Crime during the Pandemic

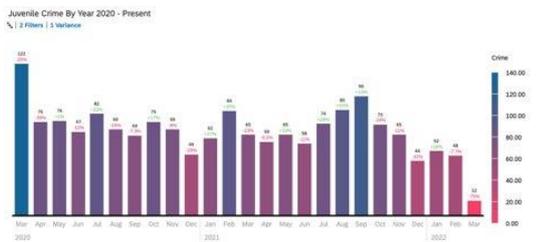

Figure 7. Juvenile Crime during the Pandemic

## 6.5 Crime Type by Area 2018-2019 (Marimekko Chart)

As evident in figure 4. Our findings resulted in a low correlation between the Crime rates before and after the COVID 19 lockdown. As initially due to the events that took place mentioned in the media regarding criminal activity such as looting, riots, and other acts of crime. Therefore, we took a different angle to explore other correlations in the types of crimes before and after the pandemic. We chose to divide the 2018-present in our dataset to see a better overview of the changes per years

Our findings in Figure 8 aligned closely with the totals in figure 3 as the physical location of Crime did not increase; rather areas with a high rate of crime (Central, 77th, and Pacific) saw a reshuffling where crime increased in these spots along with the type of criminal activity in those areas. As seen in Figure 8 if we look at the types of crimes for 2018-2019 (before the pandemic) there is a large spread in types of crime with the largest section being Battery and assault, items stolen from vehicles and vehicles stoles (Figure 9) with an overall total amount of Crime of 446,809 crime from 2018-2019. This served as an out based line for the measure as Crime type seems consistent with what is expected to be within a dense city population.

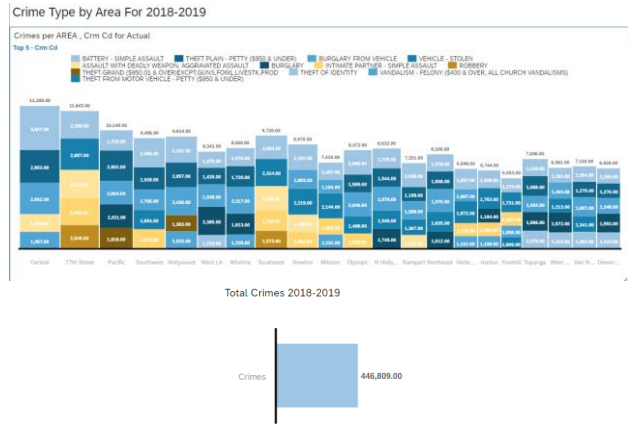

Figure 8. Crime Type by Area For 2018-2019

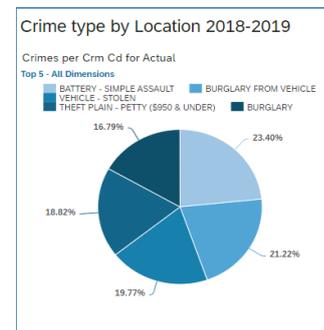

Figure 9. Crime Type by Area For 2018-2019

## 6.6 Crime Type by Area 2020-Present (Marimekko Chart)

When we compared 2020-Present at 442K cases versus 446K cases in 2018-2019 we saw a slight decrease in total crime rates. We interpreted this information as Crime has a low relationship when compared to the pandemic as a factor that influences crime based on the amount of COVID 19 cases reported and total crime. We did see a relationship between the different types of Crime 2018-2019 change compared to 2020-present. During the pandemic, there was a strong emphasis on social distance and remaining at home due to lockdowns. With our model, this explains why there was a shift in crime type because in figure 11 the types of Crime broken by top 5 based on crime description into percentages reflects that most of the crimes committed were "vehicles stolen" over "battery and assault" which was higher in 2018-2019. When taken into consideration the locations of crimes are reflected in the Geomap (Figure 3) many of these areas such as 77th and Central are places with dense communities relying on public transit or public parking if they own a vehicle. It is a fair assessment to determine the emphasis on social distancing and staying/working at home deterred people from public spaces (i.e., going out) while increasing the time they spent away from their vehicles; This potentially explains why the total in accounts of crime did not see a significant change. Rather criminals aimed for something else as a response to the situation. There is some correlation when looking at types of crime in 2018-2019 versus 2020- present. Indicated by figure

11 the percent of vehicles stolen overtook battery and assault the following year compared to 2018-2019 (figure 9).

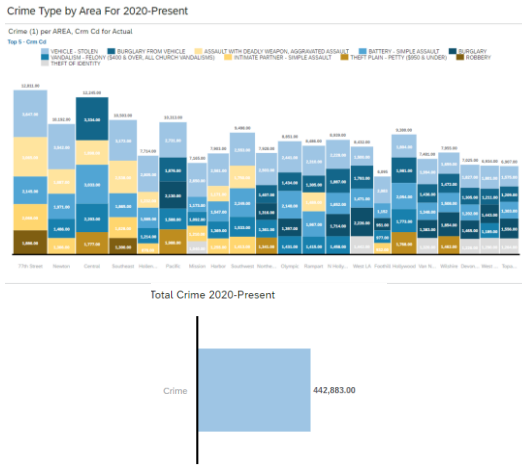

Figure 10. Crime Type by Area For 2020-Present

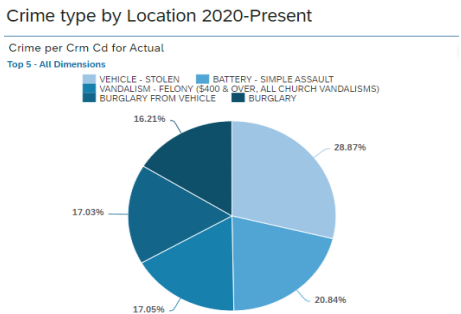

Figure 11. Crime Type by Area For 2020-Present

**6.7 Regression analysis Crime Rates from 2018-2019**

3 targets and applied a regression model, the first 3 as shown in figure 12 show high levels of correlation between Crime compared against (location, age, and type) (excluding DR_NO, DateRptd, Rtp Dist No,Part1-2,Crm Cd 1, Crm Cd 2, LAT_CRM, LON_CRM, LAT_DIV,LON_DIV,CM CD DESC Mocodes, Premis Desc, Crm Cd 3, Crm Cd 4, DIV_CITY, DIV ADDRESS) meaning there is strong evidence in the influence of these factors contributing to crime Meaning there are more influences outside of the data that impact the relationship between the target crime and the areas; As well as showing crimes do not discriminate against factors such as victim descendant or victim sex. The is a similar store when you look at the Regression model results from 2020-Present (figure 14) as 1 of the models show a similar trend when looking at the prediction confidence and root mean squared. The correlation between regression within the age group of crimes most notably victim sex and victim age in Model 2 (figure 12 and figure 13). The reason for Victim parameters is when comparing strictly at the crime age the results provided as inconclusive as labels for Crime age are Juvenile crimes and adult crime with more crimes being committed by adult this creates a scenario where the regression assumes the older an individual becomes the greater the likelihood, they will commit a crime.

This makes a bad model as it does not create a good representation since the influence on age cannot be changed. Instead, we looked at victim age excluding crm cd1,2,3,4, rpt dist No as the regression offers a better window (figure 13 and figure 15) in parameters criminals can target with age, descent, and sex being the top contenders; with sex jumping from 10.59% in 2018-2019 to 25.07% in 2020-Present meaning the sex of a victim was more a factor look at crime based on the targets age. While this does not offer a directly one to one relationship parameters such as quantity of product sold compared to revenue earned or income compared to education level, it does give a good indication of crime within the area and what factors tend to influence the decision of crime.

Figure 12. Regression For Crime compared to type 2018-2019

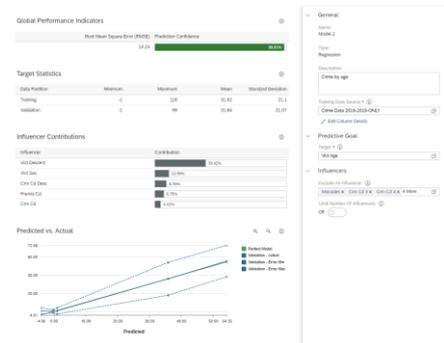

Figure 13. Regression Target Victim Age 2018-2019

Figure 14. Regression For Crime compared to type from 2020-Present

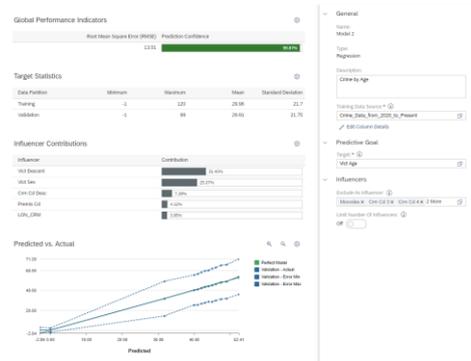

Figure 15. Regression For Crime Target Vict Age 2020-Present

**6.8 Examination of Juvenile Arrest Rates by Area 2019-2020**

While the overall number of juvenile arrests declined, a different reality from the overall crime rates, we were also interested in the areas that the crimes were taking place- Were serious crimes happening in the same neighborhoods as social conditions changed, or did neighborhoods that were high crime prior to the pandemic also the areas of highest crime in pandemic conditions?

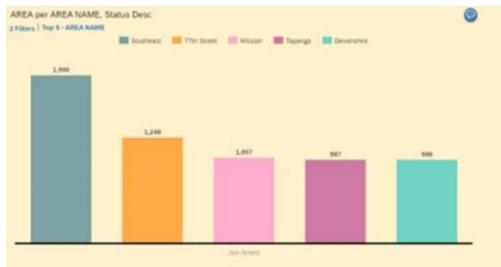

Figure 16 Crime Areas in 2019 and 2020

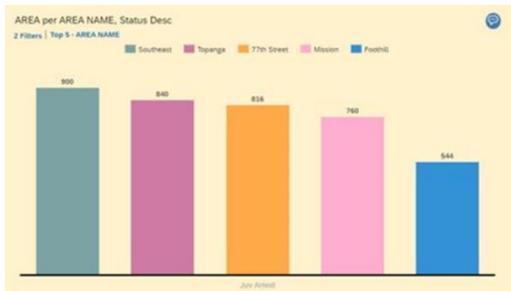

Figure 17 Crime Areas after 2020

By ranking the areas where the most arrests occurred, as seen in Figures 16 & 17 we can see a shift in 4 of the top 5 areas, and an unexpected neighborhood entered the top 5. Access to this information could be useful to police departments as they plan out their patrols.

## 6.9 Analysis of weapons used in juvenile arrests 2019-2020

In addition to knowing the areas where crimes occurred, it might be useful to analyze the weaponry used in each arrest that was committed by juveniles. In both 2019 and 2020 the number one crime involving weapons was striking with hands and fists, so that was filtered out of the analysis. Additionally, any crime involving an unknown weapon of any kind was also filtered out.

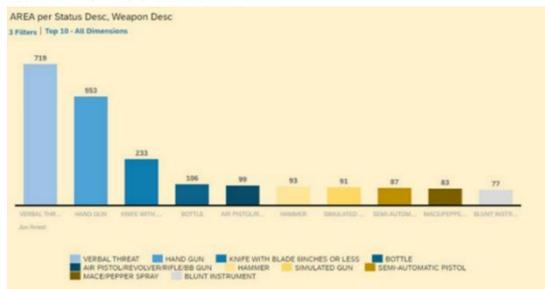

Figure 18. 2019 Juvenile arrests involving weapons by description

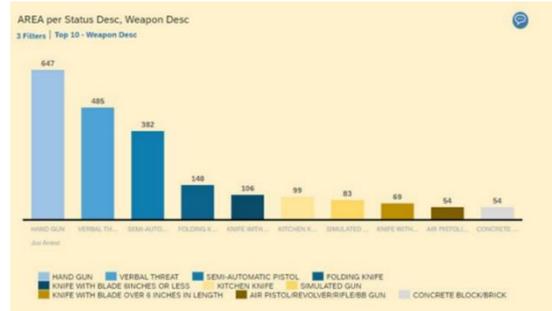

Figure 19. 2020 Juvenile Arrests involving weapons by description

In Figures 18 & 19 we can see in 2019 there was a variety of weapons used, while in 2020 the weapons used were more homogeneous. Four of the top 10 weapons used in 2020 were knives. The use of concrete blocks and bricks were also included in the top ten. This information could be useful to law enforcement in preparing for encounters in the field.

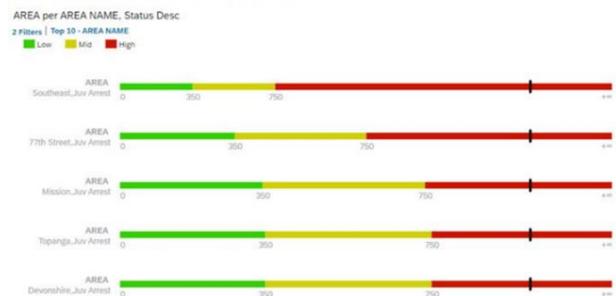

Figure 20. Juvenile arrests in 2019

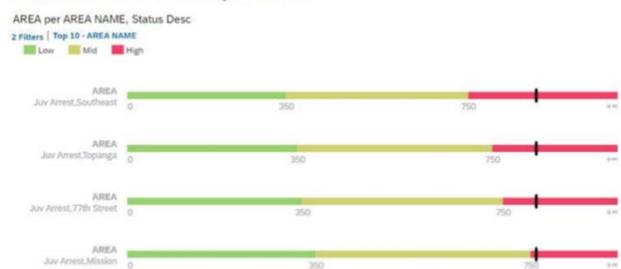

Figure 21. Juvenile arrests in 2020

## 6.10 Threshold Analysis of Juvenile Arrests by Area 2019-2020

The final component to our analysis of juvenile crime before and during the pandemic involves a threshold analysis of arrests by area. Arrests were divided into 3 categories: Low, Medium, and High. Low was set at 0-350 arrests, mid included 350-750 arrests, and areas in the high category had more than 750 arrests. Police departments might use this information to schedule resources more effectively in the neighborhoods they serve. As can be observed in Figures 20

& 21, we can also see there were fewer neighborhoods with a high number of juvenile arrests in 2020 than in 2019.

## 7. Conclusion

We found that there was very little relationship between crime and the pandemic. Crime rates dropped during the pandemic by roughly 10% meaning crime rate tends to be independent despite a global event. Type of crimes shows a variation; less people were outside meaning (1) less assault, (2) less battery stolen, and (3) more vehicles stolen presumably due to people not driving as much. Age has a plausible relationship, but inherently limited by factors such schooling, demographic, area, and specific age group. Criminals seem to find a way to make it happen.